\documentclass[prl,twocolumn,superscriptaddress,showpacs,amsmath,amssymb]{revtex4}
\usepackage{graphicx}
\usepackage{dcolumn}
\usepackage{bm}
\usepackage{color}

\newcommand{\beq}{\begin{eqnarray}}
\newcommand{\eeq}{\end{eqnarray}}

\begin{document}

\title{
Symmetry Protected Topological Order and Spin Susceptibility in Superfluid $^3$He-B
}

\author{Takeshi Mizushima}
\email{mizushima@mp.okayama-u.ac.jp}
\affiliation{Department of Physics, Okayama University,
Okayama 700-8530, Japan}
\author{Masatoshi Sato}
\email{msato@nuap.nagoya-u.ac.jp}
\affiliation{The Institute for Solid State Physics, The University of Tokyo, Chiba, 277-8581, Japan}
\author{Kazushige Machida}
\affiliation{Department of Physics, Okayama University,
Okayama 700-8530, Japan}
\date{\today}

\begin{abstract}

We here demonstrate that the superfluid $^3$He-B under a magnetic field
 in a particular direction stays
 topological due to a discrete symmetry, that is, in a symmetry
 protected topological order. 
Due to  the symmetry protected topological order, 
helical surface
 Majorana fermions in the B phase remain gapless and their Ising spin character
 persists.
We unveil that the competition between the Zeeman magnetic field and
 dipole interaction involves anomalous quantum phase transition where
 topological phase transition takes place together with spontaneous
 breaking  of symmetry.
Based on the quasiclassical theory, we illustrate that the phase
 transition is accompanied by 
anisotropic quantum criticality of spin susceptibilities on the
 surface, which is detectable in NMR experiments. 

\end{abstract}



\maketitle

{\it Introduction.---}
Superfluid $^3$He-B is one of the most concrete
examples of time-reversal invariant topological superfluids (TSFs)~\cite{review1, review2}, where
the ground state wave function supports a nontrivial bulk
topological invariant in three spatial dimensions~\cite{qi, schnyder,
sato09, sato10, kitaev,volovik2009v1}.
As a consequence of the bulk-edge correspondence,
helical
Majorana fermions live on its specular surface, and their self conjugate
property gives rise
to an Ising-like anisotropy of spin susceptibility~\cite{stone,chung,nagato,volovik2009,shindou,tsutsumi,TM,silaev}.
Recently the surface Majorana cone has been detected in
experiments~\cite{murakawa}.

Since the topological superfluidity in $^3$He-B, which is
categorized to class DIII~\cite{schnyder},  is ensured
by time-reversal invariance, it is sometimes stated that any
time-reversal breaking such as a finite magnetic field immediately wipes
out the topological nature.
Indeed, in the presence of a strong magnetic field,
the Majorana Ising spin ceases to exist~\cite{volovik2010}. 
However, as argued in the Letter, a more careful consideration on the
basis of microscopic calculations and symmetry of the system points to a
different conclusion. 
It is worth mentioning that 
the robustness of a topological phase transition for a
topological insulator against the time-reversal breaking 
is proposed in an extended Kane-Mele model.~\cite{altman}.

\begin{figure}[b!]
\includegraphics[width=85mm]{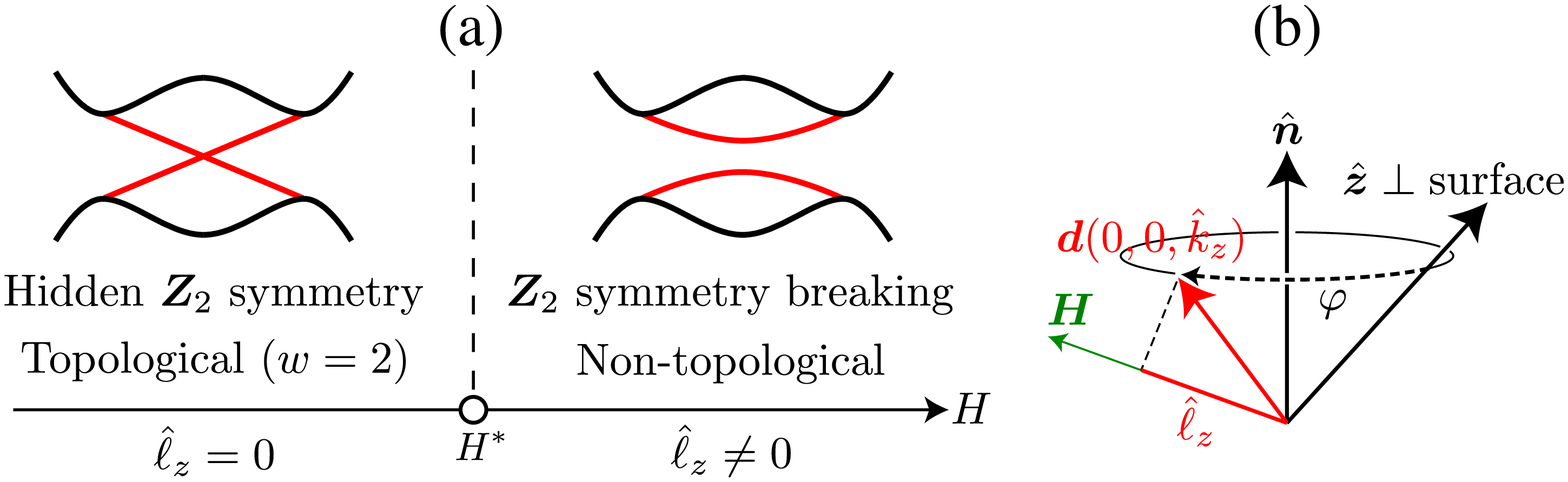}
\caption{(Color online) (a) Schematic phase diagram of $^3$He-B under a parallel magnetic field, where $H^{\ast}$ involves the topological phase transition with spontaneous symmetry breaking. (b) Relation between $\hat{\ell}_z$ and the the orientation of ${\bm d}(0,0,\hat{k}_z)$ for an arbitrary $(\hat{\bm n},\varphi)$ at the surface.}
\label{fig:phase}
\end{figure}

In this Letter, we show that $^3$He-B under a magnetic field in a particular
direction stays topological as a symmetry protected topological order.  
In spite of the time-reversal breaking due to the magnetic field,  the
topological property is retained by a hidden ${\bm Z}_2$ symmetry
that is obtained by a combination of 
time-reversal and an ${\rm SO}(3)_{{\bm L}+{\bm S}}$ rotation. Because
this ${\bm Z}_2$ symmetry restores a chiral symmetry of the
microscopic Hamiltonian, helical surface Majorana fermions in the
B-phase remain to be gapless, and the Ising spin character of the
Majorana
fermions persists unless the discrete symmetry is spontaneously
broken.
Finally, we come
to the conclusion that at a critical Zeeman field $H^{\ast}$ 
the system undergoes anomalous quantum phase transition
in the sense that topological
phase transition takes place together with spontaneous breaking of
the ${\bm Z}_2$ symmetry (Fig.~\ref{fig:phase}(a)). 

The phase transition is described by a pair of an order parameter
$\hat{\ell}_z$ and a topological number $w$, which are defined 
in Fig.~\ref{fig:phase}(b) and in the text below Eq.~(\ref{chiral}). 
Conventionally, topological phase transitions are accompanied by
creation or destruction
of gapless surface states and gap closing of
bulk quasi-particle spectra.
While the anomalous quantum phase transition in the above involves
the destruction of gapless surface states, it does not show the bulk gap
closing. 
Instead, there appears a long range order due to symmetry breaking.
Moreover, 
it takes place between two conceptional different quantum orders:
At a critical magnetic field, 
the system undergoes a transition from topologically
ordered state ($w\neq 0$) to conventionally ordered one ($\hat{\ell}_z\neq 0$).

The symmetry protected topological order is closely associated with the order parameter manifold
of $^3$He-B.
Ignoring the dipole interaction and a Zeeman field, the bulk $^3$He-B
spontaneously reduces the symmetry ${\rm SO}(3)_{\bm L}\!\times\!{\rm
SO}(3)_{\bm S}\!\times\!{\rm U}(1)$ to ${\rm SO}(3)_{{\bm L}+{\bm
S}}$~\cite{vollhardt}. The gap function is $\Delta (\hat{\bm
k},{\bm r}) \!=\! i{\bm \sigma}\cdot{\bm d}(\hat{\bm
k},{\bm r})\sigma _y$, where $d_{\mu}\!=\!d_{\mu\nu}\hat{k}_{\nu}$ and,
\beq
d_{\mu\nu}({\bm r}) = e^{i\vartheta}R_{\mu\nu}(\hat{\bm n},\varphi)
\Delta _{\nu}({\bm r}). 
\label{eq:dmn}
\eeq
The broken symmetry ${\rm SO}(3)_{{\bm L}-{\bm S}}$, the relative
rotation between spin and orbital
spaces, is described by $R_{\mu\eta}(\hat{\bm n},\varphi)$ with the
rotation axis $\hat{\bm n}$ and the angle $\varphi$. 
Here the repeated Greek indices imply the sum ($\mu,\nu,\eta
\!=\! x,y,z$) and $\sigma _{\mu}$ denotes the Pauli matrices in the spin
space. In general, the dipole interaction
acting as a small perturbation chooses a particular state of $(\hat{\bm
n},\varphi)$.

To quantitatively determine $\hat{\ell}_z(\hat{\bm
n},\varphi)$ and $H^{\ast}$, we here utilize the quasiclassical theory
which takes account of dipole interaction and Zeeman energy on equal
footing. In a slab geometry, the finite $H^{\ast}$ results from the competition between 
dipole and magnetic energies, where the former (latter) favors
the symmetry protected topological (a non-topological) order.
It is found that since the topological order protects the Majorana Ising
spins, the phase transition is accompanied by anomalous critical
behaviors of spin susceptibilities on the surface.

{\it Surface bound states.---}
Let us start with the mean-field Hamiltonian density in the Nambu representation,
\beq
\underline{\mathcal{H}}({\bm r}_1,{\bm r}_2) = \left[
\begin{array}{cc}
\epsilon ({\bm r}_1,{\bm r}_2) & \Delta ({\bm r}_1,{\bm r}_2) \\
-\Delta^{\ast}({\bm r}_1,{\bm r}_2) & -\epsilon^{\ast}({\bm r}_1,{\bm r}_2) 
\end{array}
\right] + \underline{V_{\rm Z}}\delta({\bm r}_{12}).
\label{eq:hami}
\eeq
In this paper, we set $\hbar \!=\! k_{\rm B} \!=\! 1$. 
Equation (\ref{eq:hami}) consists of 
$\epsilon ({\bm r}_1,{\bm r}_2)\!=\!\delta ({\bm r}_{12}) (-{\bm
\nabla}^2/2M-E_{\rm F})$ and the Zeeman energy $\underline{V_{\rm Z}} \!\equiv\!
-\mu _{\rm n}H_{\mu}{\rm diag}(\sigma _{\mu},-\sigma^{\ast}_{\mu})$,
where $M$, $E_{\rm F}\!=\!k^2_{\rm F}/2M$, and $\mu _{\rm n}$ are the
mass, Fermi energy, and magnetic moment of $^3$He atoms. The pair potential for $^3$He-B, $\Delta
({\bm k},{\bm r}) \!\equiv\! \int d{\bm r}_{12}e^{-i{\bm k}\cdot{\bm
r}_{12}} \Delta ({\bm r}_1,{\bm r}_2)$ with Eq.~(\ref{eq:dmn}), is simplified to 
$\Delta ({\bm k},{\bm r}) \!=\! U(\hat{\bm n},\varphi)\Delta _0({\bm k},{\bm r})U^{\rm T}(\hat{\bm n},\varphi)$ 
with $U(\hat{\bm n},\varphi)\!\in\!{\rm SU}(2)$ and $\Delta _0({\bm k},{\bm r}) \!=\! i\sigma _{\mu}\sigma _y \Delta_{\mu}({\bm r})\hat{k}_{\mu}$.


We first diagonalize Eq.~(\ref{eq:hami}) as $\int d{\bm r}_2 \underline{\mathcal{H}}({\bm r}_1,{\bm r}_2){\bm \varphi}_E({\bm r}_2) \!=\! E{\bm \varphi}_E({\bm r}_1)$, that is, the Bogoliubov-de Gennes (BdG) equation, where $E$ and ${\bm \varphi}_E$ describe the energy and wavefunction of quasiparticles. This is solved within the Andreev approximation ${\bm \nabla}^2 \!\rightarrow\! i {\bm v}_{\rm F}\cdot{\bm \nabla}$ and the uniform pair potential $\Delta _{\mu}({\bm r})\!=\! \Delta _0$, where ${\bm v}_{\rm F} \!=\! \hat{\bm k}v_{\rm F}$ is the Fermi velocity. In this work, we consider the B-phase sandwiched by two specular walls which are normal to the $\hat{\bm z}$-axis. 
For ${\bm H} \!=\! {\bm 0}$, the dispersion of the surface Andreev bound state
(SABS) is given by $E_0 ({\bm k}_{\parallel}) \!=\! \pm \frac{\Delta
_0}{k_{\rm F}} | {\bm k}_{\parallel}|$ with ${\bm k}_{\parallel}$ being
the momentum in the $xy$-plane~\cite{chung,nagato,volovik2009v1}. The
corresponding wave functions are expressed as ${\bm
\varphi}^{(\pm)}_{0,{\bm k}_{\parallel}}({\bm r}) \!\propto\! e^{i{\bm
k}_{\parallel}\cdot{\bm r}_{\parallel}}e^{-z/\xi}\sin(\sqrt{k^2_{\rm
F}-k^2_{\parallel}}z)\underline{\mathcal U}(\hat{\bm n},\varphi){\bm
\Phi}^{(\pm)}_{\bm k}$, where $(\pm)$ correspond to the positive and
negative energy states and we set $\underline{\mathcal{U}}\!\equiv\!
{\rm diag}(U,U^{\ast})$. We also introduce ${\bm \Phi}^{(+)}_{\bm k}
\!=\! \underline{\tau}_x {\bm \Phi}^{(-)\ast}_{{\bm
k}_{\parallel}}\!\equiv\! (1, -ie^{i\phi _{\bm k}},-e^{i\phi_{\bm
k}},-i)^{\rm T}$ with $\phi _{\bm k} \!=\!
\tan^{-1}(\hat{k}_y/\hat{k}_x)$ and $\underline{\tau}_{\mu}$ being the
Pauli matrices in the Nambu space.  

For a finite ${\bm H}$, the dispersion of the SABS is obtained from the linear combination, ${\bm \varphi}_{{\bm k}_{\parallel}} \!=\! a_+ {\bm \varphi}^{(+)}_{0,{\bm k}_{\parallel}} \!+\! a_- {\bm \varphi}^{(-)}_{0,{\bm k}_{\parallel}} $, as 
\beq
E ({\bm k}_{\parallel}) = \pm \sqrt{[ E_0 ({\bm k}_{\parallel})]^2 + [\mu _{\rm n}H\hat{\ell}_{z}(\hat{\bm n},\varphi)]^2},
\label{eq:Esabs}
\eeq
where $\hat{\ell}_{\nu}(\hat{\bm n},\varphi)\!\equiv\!\hat{h}_{\mu}R_{\mu {\nu}}(\hat{\bm n},\varphi) $
with $\hat{h}_{\mu} \!=\! H_{\mu}/H$.

{\it Symmetry protected topological phase ---}
As we showed in Eq.~(\ref{eq:Esabs}), if $\hat{\ell}_z\!=\!0$,
the SABS remains gapless even in the presence of a magnetic field.
From topological point of view, however, this seems to be a puzzle:
Because the magnetic field breaks the time-reversal invariance, 
topological protection as a time-reversal invariant TSF does not work
any more.
Nevertheless, no gap opens in the SABS if the
magnetic field satisfies $\hat{\ell}_z\!=\!0$. 

First, one should notice that
$\hat{{\bm \ell}}$ 
itself
could be affected by a Zeeman magnetic field.
Therefore, an immediate solution for this puzzle might be
that if one applies a magnetic field, $\hat{{\bm \ell}}$ changes so as
$\hat{\ell}_z \!\neq\! 0$.
However, as is shown below, if the Zeeman field is parallel to
the $xy$-plane (say, along $\hat{\bm x}$), this is not the case:
There exists a symmetry that ensures $\hat{\ell}_z\!=\!0$.
Interestingly, we find that this symmetry
resolves the puzzle above at the same time, by providing another topological
protection of the SABS.

Let us first consider symmetry of the system.
Among SO(3)$_{{\bm L}+{\bm S}}$ rotations under which 
microscopic interactions of ${}^3$He atoms are invariant, 
the slab geometry considered here preserves its subgroup
%
SO(2)$_{{\bm
L}+{\bm S}}$ rotation $U(\theta)$ in the $xy$-plane.
The point is that while the Zeeman term along $\hat{\bm x}$ explicitly breaks both of the
time-reversal symmetry and the SO(2)$_{{\bm L}+{\bm S}}$ rotation
symmetry above,  it does not break a combination of them.
In fact, in this case, the flipped magnetic field  by
time-reversal ${\cal T}$ is recovered by the $\pi$-rotation in the
$xy$-plane.
Therefore, the microscopic Hamiltonian of ${}^3$He atoms is invariant
under the discrete symmetry given by ${\cal T}U(\pi)$.
We notice here that $\hat{\ell}_z$ mentioned above is transformed
nontrivially as $\hat{\ell}_z \rightarrow
-\hat{\ell}_z$ under this symmetry. 
Therefore $\hat{\ell}_z$ is an order parameter of the discrete symmetry,
and it should be zero unless the discrete symmetry is spontaneously broken.

Remarkably, one can introduce a topological invariant if the discrete symmetry is not spontaneously broken.
In that case, the BdG Hamiltonian (\ref{eq:hami}) in the momentum space is manifestly invariant under the discrete symmetry,
$\underline{\cal H}(k_x, k_y, -k_z) \!=\! {\cal T} U(\pi)\underline{\cal H}(k_x,k_y,k_z)U^{-1}(\pi){\cal T}^{-1}$, 
where ${\cal T}\!=\!i\sigma_y K$ is time-reversal with complex conjugate
operator $K$ and
$U(\pi)\!=\!i\sigma_z\tau_z$ is the $\pi$-rotation.  
Therefore, 
combining it with the particle-hole symmetry of the BdG Hamiltonian, $\underline{\cal C}\underline{\cal H}({\bm k})\underline{\cal
C}^{\dagger}\!=\!-\underline{\cal H}^{*}(-{\bm k})$ with $\underline{\cal C} \!=\! \underline{\tau}_x K$,
one obtains the relation
$
\underline{\Gamma} \underline{\cal H}(k_x,k_y,k_z)\underline{\Gamma}^{-1}\!=\!-\underline{\cal H}(-k_x,-k_y, k_z)
$
with $\underline{\Gamma} \!=\! \sigma_x\underline{\tau}_y$.
On the $k_z$ axis, this reduces to the so-called
chiral symmetry
\begin{eqnarray}
\{\underline{\Gamma}, \underline{\cal H}(0,0,k_z)\}=0.
\label{chiral}
\end{eqnarray}
Thus, following Refs.~\cite{sato2009,sato2011}, one can introduce the
following one-dimensional (1D) winding number 
$w\!=\!-\frac{1}{4\pi i}\int_{-\infty}^{\infty}dk_z\left. 
{\rm tr}[\underline{\Gamma}\underline{\cal H}^{-1}\partial_{k_z}\underline{\cal H}]
\right|_{{\bm k}_{\parallel}\!=\!{\bm 0}}$,
which are evaluated as $w \!=\! 2$ for $\mu_{\rm
n}H \! <\! E_{\rm F}$ ($\Delta_{\perp}>0$). 
Therefore, the system is
topologically non-trivial, and the bulk-edge correspondence implies that
the SABS satisfies $E({\bm k}_{\parallel}) \!=\! 0$ at ${\bm
k}_{\parallel}\!=\! {\bm 0}$ even in the presence of the magnetic field.
It should be noted here that one needs the discrete symmetry specific to
this system in order to define $w$.
Therefore a topological phase realized here is a symmetry
protected topological order \cite{wen,pollmann}.

{\it Majorana Ising Spin.---}
In the absence of the magnetic field, 
it have been known that helical Majorana fermions in $^3$He-B 
have Ising-like spin
density\cite{stone,chung,nagato,volovik2009,shindou,tsutsumi,TM,silaev}.
Now we show that the symmetry
protected topological order discussed above retains the
Ising spin character of Majorana fermions.

To see this, we use a general symmetry property of the SABS. 
Let us consider the low energy limit where only the
zero energy SABSs at ${\bm k}_{\parallel}\!=\!{\bm 0}$ contribute. 
According to the index theorem of Ref.~\cite{sato2011}, Eq.~(\ref{chiral})
infers that the zero energy SABSs are
eigenstates of $\Gamma$, and the relation $w=n_--n_+$ holds on the
surface of superfluid $^3$He-B, where
$n_{\pm}$ is the number of the zero energy SABSs with the eigenvalue
$\Gamma \!=\! \pm 1$.
In the present case, 
$w \! =\! n_- \!=\! 2$ and $n_+ \!=\! 0$,   
and thus the SABS ${\bm \varphi}^{(a)}_{{\bm
k}_{\parallel}\!=\!{\bm 0}}$ satisfies $\Gamma{\bm \varphi}^{(a)}_{{\bm
k}_{\parallel}\!=\!{\bm 0}}\!=\!-{\bm \varphi}^{(a)}_{{\bm k}_{\parallel}\!=\!{\bm
0}}$ $(a\!=\!1,2)$.
Using the particle-hole symmetry, one can also put the
relation $\tau_x{\bm \varphi}^{(a)*}_{{\bm k}_{\parallel}\!=\!{\bm 0}}\!=\!{\bm
\varphi}^{(a)}_{{\bm k}_{\parallel} \!=\! {\bm 0}}$ at the same time.
From these two relations, ${\bm \varphi}^{(a)}_{{\bm k}_{\parallel} \!=\! {\bm
0}}$ has a generic form as ${\bm \varphi}^{(a)}_{{\bm
k}_{\parallel} \!=\! {\bm
0}} \!=\! [\xi^{(a)},i\xi^{(a)*},\xi^{(a)*},-i\xi^{(a)}]^{\rm T}$ with a
function $\xi^{(a)}$.
Ignoring non-zero energy modes, the
quantized field ${\bm \Psi} \!=\!
[\hat{\psi}_{\uparrow},\hat{\psi}_{\downarrow}, 
\hat{\psi}_{\uparrow}^{\dagger},\hat{\psi}_{\downarrow}^{\dagger}]^{\rm T}$
is expanded as
$\Psi(z) \!=\! \sum_{a=1,2}{\bm \varphi}^{(a)}_{{\bm k}_{\parallel} \!=\! {\bm
0}}\gamma^{(a)}$ with real $\gamma^{(a)}$, and 
from the general form of ${\bm \varphi}^{(a)}_{{\bm
k}_{\parallel} \!=\! {\bm 0}}$, one obtains
$
i\hat{\psi}_{\uparrow} \!=\! -\hat{\psi}_{\downarrow}^{\dagger}, 
$
which is a general consequence of our symmetry protected topological order.

Now following Ref.\cite{stone, chung}, one can show that the last relation,
$i\hat{\psi}_{\uparrow} \!=\! -\hat{\psi}_{\downarrow}^{\dagger}$, yields
the Ising character of the SABSs:
It is shown that among the local density operator
and the spin density operators,
which are given by
$\rho \!\equiv\! \frac{1}{2}[\hat{\psi}^{\dagger}_a\hat{\psi}_a-\hat{\psi}_a\hat{\psi}^{\dagger}_a] $ 
and $S_{\mu} \!\equiv\!
\frac{1}{4}[\hat{\psi}^{\dagger}_a(\sigma_\mu)_{ab}\hat{\psi}_b-\hat{\psi}_a(\sigma_\mu^{\rm
T})_{ab}\hat{\psi}^{\dagger}_b]$, respectively, only $S_z$ is nonzero while the
other components are identically zero.   
So, in the low energy limit, the SABSs do not
contribute to the local density fluctuation, and its local spin
density is Ising-like.
Here note that we only use a general property of the chiral symmetry, thus the Ising character is a direct consequence of our symmetry protected topological order.



{\it Quasiclassical Eilenberger theory.---}
Let us now microscopically determine $\hat{\ell}_z$ and $H^{\ast}$. For this purpose, we here utilize the quasiclassical Eilenberger theory, which provides a quantitative theory for superfluid $^3$He at low pressures~\cite{serene}. 
This is based on the quasiclassical Green's functions $\underline{g} \!\equiv\! \underline{g}(\hat{\bm k},{\bm r};i\omega _n)$ with the Matsubara frequency $\omega _n \!=\! (2n+1)\pi T$ ($n\!\in\! \mathbb{Z}$) and the $2\!\times\! 2$ unit matrix $\sigma _0$
\beq
\underline{g} = \left[
\begin{array}{cc}
{\sigma}_{0} g_{0} + {\sigma}_{\mu} g_{\mu}   & i\sigma _y f_0 + i {\sigma}_{\mu} {\sigma}_y f_{\mu}  \\ 
i\sigma _y f^{\dag}_0 +i \sigma _y {\sigma}_{\mu}{f}^{\dag}_{\mu}  & {\sigma}_{0} g^{\dag}_{0} + {\sigma}^{\ast}_{\mu} g^{\dag}_{\mu}
\end{array}
\right].
\eeq
The evolution is governed by the Eilenberger equation
$
[i\omega _n \underline{\tau}_z 
- \underline{\mathcal{S}}(\hat{\bm k},{\bm r}), 
\underline{g}
] \!=\!- i {\bm v}_{\rm F} \!\cdot{\bm \nabla}\!
\underline{g}$.
The $4\!\times\! 4$ matrix $\underline{\mathcal{S}}$ consists of the Zeeman energy $\underline{V_{\rm Z}}$ and the self-energies,
\beq
\underline{\mathcal{S}}(\hat{\bm k},{\bm r}) = \frac{1}{1+F^{\rm a}_0}\underline{\tau}_z\underline{V_{\rm Z}} + \left[
\begin{array}{cc}
\nu _{\mu}(\hat{\bm k},{\bm r}){\sigma}_{\mu} & \Delta(\hat{\bm k},{\bm r}) \\
\Delta^{\dag}(-\hat{\bm k},{\bm r}) & 
{\nu}^{\ast}_{\mu}(\hat{\bm k},{\bm r}){\sigma}^{\ast}_{\mu}
\end{array}
\right],
\label{eq:s}
\eeq
where $\nu_{\mu}$ denotes the Fermi liquid corrections obtained as
$
\nu _{\mu} (\hat{\bm k},{\bm r}) \!=\! \sum _{\ell}
A^{\rm a}_{\ell}\langle P_{\ell}(\hat{\bm k}, \hat{\bm k}^{\prime}) g_{\mu}(\hat{\bm k},{\bm r};i\omega _n)
\rangle _{\hat{\bm k}^{\prime},\omega_n}$. $P_{\ell}$ is the Legendre polynomials and $\langle \cdots\rangle
_{\hat{\bm k},\omega_n} \!=\! T\sum _{|\omega _n|<E _{\rm c}}\int
\frac{d\hat{\bm k}}{4\pi}$ with a cutoff $E_{\rm c}$. The coefficient $A^{\rm
a}_{\ell}\!\equiv\!F^{\rm a}_{\ell}/(1+F^{{\rm a}}_{\ell}/(2\ell+1))$ is
parameterized with the antisymmetric Fermi liquid parameters, $F^{\rm
a}_0 \!=\! -0.695$ and $F^{\rm a}_1 \!=\! -0.5$~\cite{vollhardt}. 

The gap equation is obtained as 
$
\Delta _{ab}(\hat{\bm k},{\bm r}) \!=\! \langle V^{cd}_{ab}(\hat{\bm k}, \hat{\bm k}^{\prime}) [i\sigma _{\mu}\sigma _y f_{\mu}(\hat{\bm k}^{\prime},{\bm r};i\omega _n)]_{cd}
\rangle _{\hat{\bm k}^{\prime},\omega_n}
$,
with $a,b,c,d \!=\! \uparrow,\downarrow$. At the low pressure limit, the pair interaction of $^3$He atoms is described as 
$
V^{cd}_{ab}(\hat{\bm k}, \hat{\bm k}^{\prime})
\!=\! 3 |\Lambda| \hat{k}_{\mu}\hat{k}^{\prime}_{\mu} \delta _{ac} \delta_{bd} 
- Q_{\mu\nu}(\hat{\bm k},\hat{\bm k}^{\prime})(\sigma_{\mu})_{ac}(\sigma_{\nu})_{bd}
$.
The first term arises from the $p$-wave interaction with ${\rm SO}(3)_{\bm S}\!\times\!{\rm SO}(3)_{\bm L}\!\times\!{\rm U}(1)$ and the second term is the dipole interaction, where $Q_{\mu\nu}(\hat{\bm k},\hat{\bm k}^{\prime})$ is obtained from 
$
Q_{\mu\nu}({\bm k},{\bm k}^{\prime}) \!=\! g_{\rm D} R \int r^{-3}(\delta _{\mu\nu}\!-\!3\hat{r}_{\mu}\hat{r}_{\nu})e^{-i({\bm k}-{\bm k}^{\prime})\cdot{\bm r}}d{\bm r}
$ with ${\bm k} \!\approx\! \hat{\bm k}k_{\rm F}$. The factor $R$ includes the contributions of high energy quasiparticles~\cite{Leggett}. The dipole interaction can be expressed in terms of the partial wave series ($p$-, $f$-, and higher waves). However, since the pairing interaction between $^3$He atoms is dominated by the ${\rm SO}(3)_{\bm S}\!\times\!{\rm SO}(3)_{\bm L}\!\times\!{\rm U}(1)$ channel and the dipole interaction can be regarded as a small perturbation, we take account of only the $p$-wave contribution of $Q_{\mu\nu}(\hat{\bm k},\hat{\bm k}^{\prime})$. 
Then, the gap equation for $\Delta _{\mu} ({\bm r})$ is given by
$
\Delta _{\mu} ({\bm r})\!=\!\delta _{\mu\nu} R^{-1}_{\mu\eta}(\hat{\bm n},\varphi)
\{ ( 3|\Lambda|-\tilde{\Lambda}_{\rm D})\langle \hat{k}_{\nu}f_{\eta} \rangle  _{\hat{\bm k},\omega_n}
- 3\tilde{\Lambda}_{\rm D}[
\delta_{\eta\nu}\langle \hat{k}_{\eta}f_{\eta} \rangle _{\hat{\bm k},\omega_n}
+ \epsilon _{\eta\nu\tau}
\langle ( \hat{\bm k}\times{\bm f})_{\tau}\rangle _{\hat{\bm k},\omega_n}
] \}
$, 
with the dimensionless factor $\tilde{\Lambda}_{\rm D}\!\equiv\! \frac{3\pi}{10}g_{\rm D}R$. At the thermodynamic limit with $\Delta _x\!=\! \Delta _y \!=\! \Delta _{\parallel}$ and $\Delta _z \!=\!\Delta _{\perp}$, the gap equation reproduces $\cos\varphi \!=\! -\frac{1}{4}\frac{\Delta _{\perp}}{\Delta _{\parallel}}$~\cite{tewordt,schopohl,fishman}.

The Eilenberger and gap equations with Eq.~(\ref{eq:dmn}) provide self-consistent equations for $\underline{g}$ under a fixed $(\hat{\bm n},\varphi)$. The Eilenberger equation with $\underline{g}^2 \!=\! -\pi^2 \underline{\tau}_0$ is numerically solved with the Riccati parameterization in the system that two specular walls normal to $\hat{\bm z}$ are situated at $z \!=\! 0$ and $z \!=\! 20 \xi$. The numerical procedure is same as that in Refs.~\cite{tsutsumi,tsutsumiPRB}. We solve the gap equation with $|\Lambda|^{-1} \!=\! \pi T_{\rm c0}\sum _{|\omega _n| < E_{\rm c}}|\omega _n|^{-1}$ and $\tilde{\Lambda}_{\rm D}/\Lambda^2 \!=\! 2 \!\times\! 10^{-4}$ ($2 \!\times\! 10^{-5}$), where we set the cutoff $E_{\rm c} \!=\! 20 \pi T_{\rm c0}$. Note that since the ratio $\tilde{\Lambda}_{\rm D}/\Lambda^2$ is associated with the distortion of $\Delta _{\mu}$ in the thermodynamic limit~\cite{fishman}, it is independent of the energy cutoff. The coherence length $\xi \!\equiv\! v_{\rm F}/\pi T_{\rm c0}$ is estimated about $80 {\rm nm}$ at the zero pressure of $^3$He-B and $T_{\rm c0} \!\approx\! 1{\rm mK}$ is the critical temperature of the bulk B-phase under $H\!=\! 0$. In a slab geometry, $\hat{\bm n}$ may be assumed to be spatially uniform, because the length scale of the spatial variation of $\hat{\bm n}$ is macroscopically large~\cite{vollhardt}, compared with the thickness of the sample $20\xi \!\approx\! 1.6 \mu{\rm m}$ in typical experiments~\cite{saunders,miyawaki}. Then, the stable configuration of $(\hat{\bm n},\varphi)$ is determined by minimizing the thermodynamic potential $\delta \Omega [\underline{g}]$ whose explicit form is same as that given by Vorontsov and Sauls in Ref.~\cite{vorontsov}.
Note that for the thickness $20 \xi$, the magnetic field induces the first-order phase transition from B-phase to A- or planar phase at the critical field $H _{\rm AB} \!=\! 0.09 \pi T_{\rm c0}/\mu _{\rm n} \!\approx\! 3.6 {\rm kG}$~\cite{TMfull}.

\begin{figure}[t!]
\includegraphics[width=40mm]{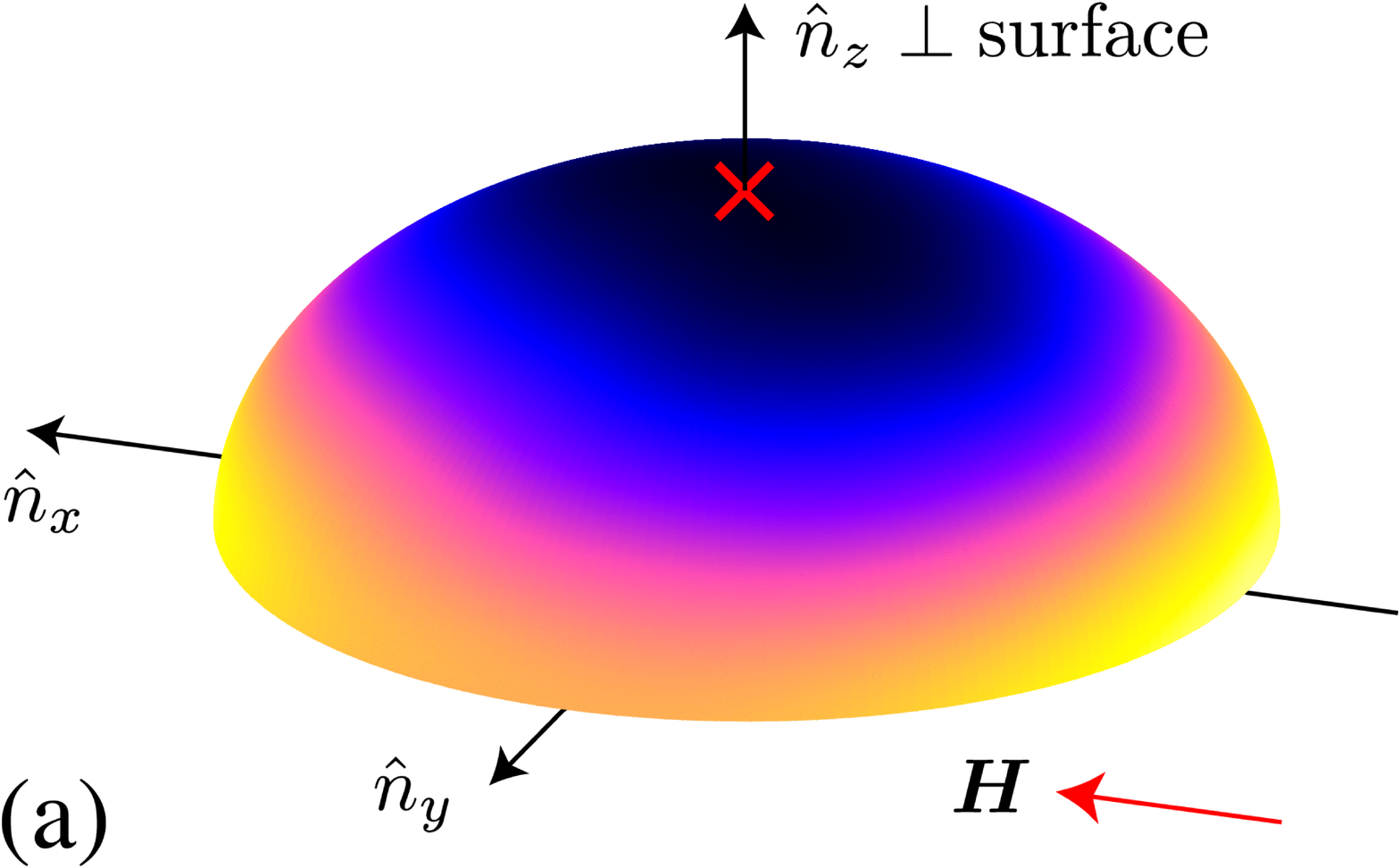}
\includegraphics[width=40mm]{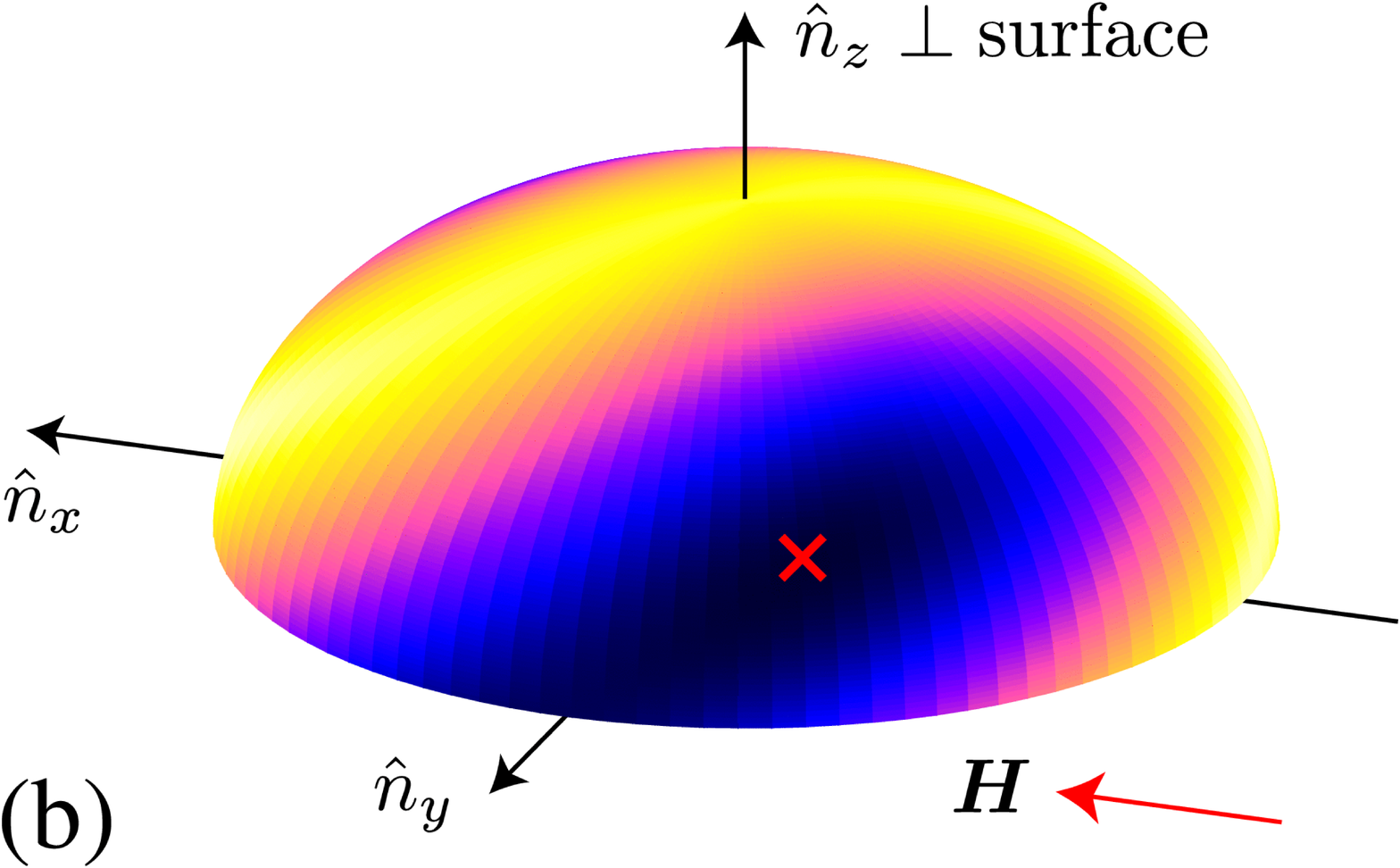}
\includegraphics[width=40mm]{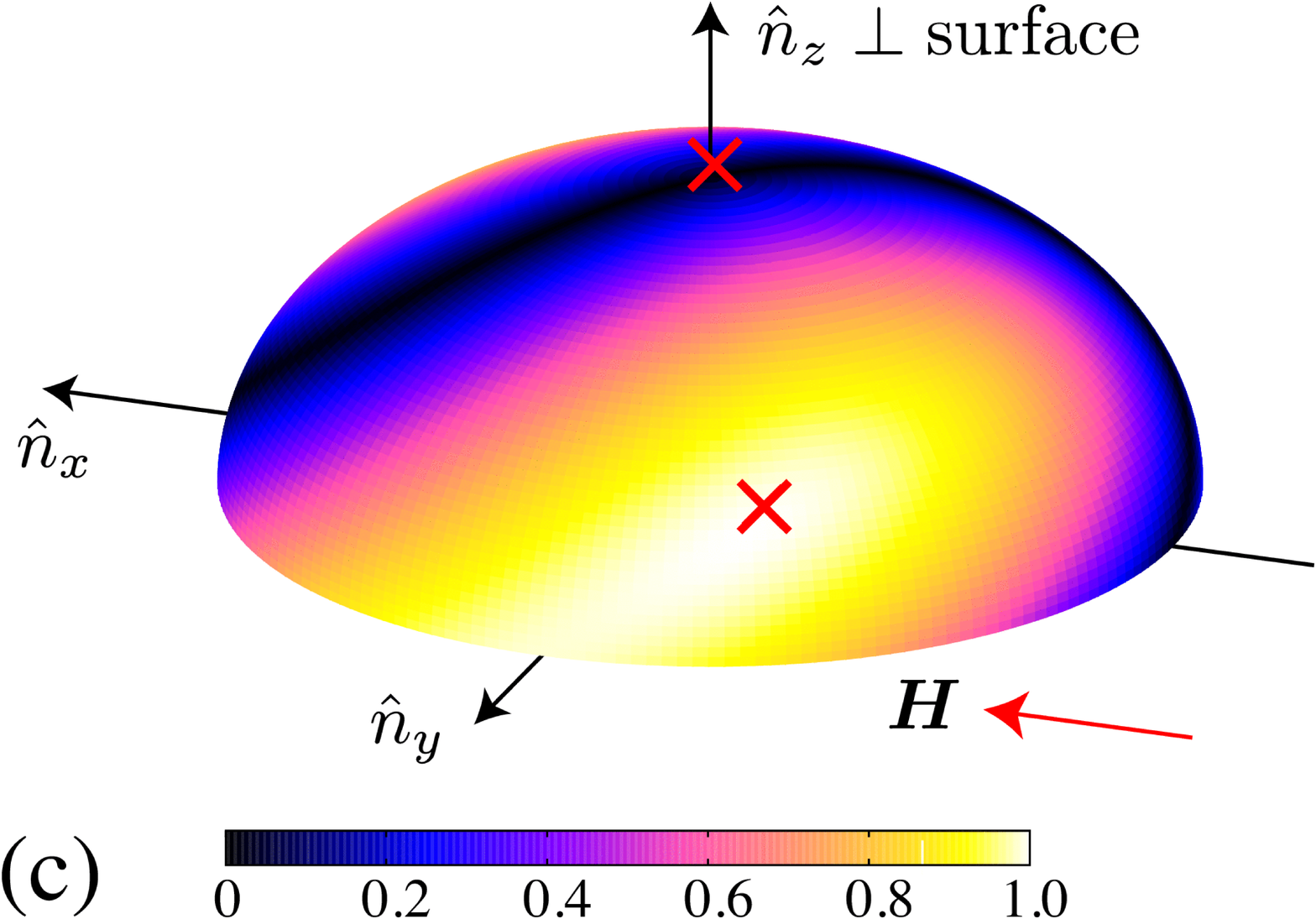}
\includegraphics[width=40mm]{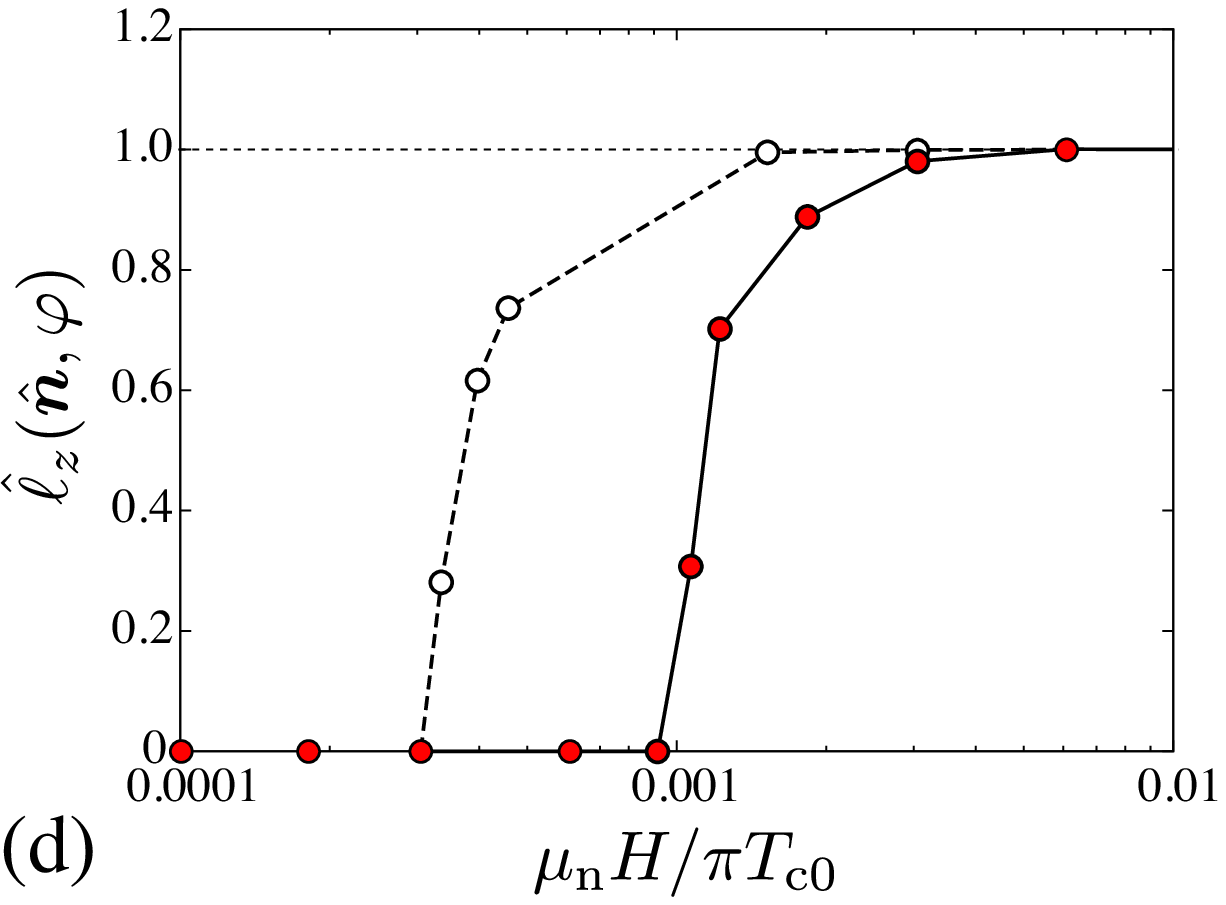}
\caption{(Color online) Energy landscape on the unit sphere of $\hat{\bm n}$, $\delta\Omega (\hat{\bm n})$, at $\mu _{\rm n}H/\pi T_{\rm c0}\!=\! 9.2\!\times\!10^{-4}$ (a) and $0.0061$ (b) where we fix $\varphi/\pi \!=\! - 0.5537$ which minimizes the dipole interaction. We also set ${\bm H} \!\parallel\! \hat{\bm x}$ and $T/T_{\rm c0} \!=\! 0.2$. The bright (dark) color depicts the higher (lower) energy. The energy gap $\min|E({\bm k}_{\parallel})|$ of Eq.~(\ref{eq:Esabs}) is displayed in (c). (d) Field dependence of $\hat{\ell}_z$ estimated with the stable $(\hat{\bm n},\varphi)$ for $\tilde{\Lambda}_{\rm D}/\Lambda^2 \!=\! 2 \!\times\! 10^{-4}$ (the solid line) and $2 \!\times\! 10^{-5}$ (the dashed line).}
\label{fig:n}
\end{figure}

{\it Numerical results.---}
Since the $\hat{\bm n}$-vector always points to $\hat{\bm z}$ in the case of the perpendicular field ${\bm H}\!\parallel\!\hat{\bm z}$, we here consider a parallel field ${\bm H}\!\parallel\!\hat{\bm x}$. Figures~\ref{fig:n}(a) and \ref{fig:n}(b) describe the energy landscape on the unit sphere of $\hat{\bm n}$. Figure~\ref{fig:n}(c) depicts the energy gap of the SABS evaluated from Eq.~(\ref{eq:Esabs}). The SABS becomes gapless along the certain trajectory on the sphere of $\hat{\bm n}$, which coincides with the condition of $\hat{\ell}_z\!=\!0$. The stable configuration of $\hat{\bm n}$ is determined as a consequence of the interplay between dipole interaction and Zeeman energy. The former favors the situation where $\hat{\bm n}$ is normal to the surface, namely, $\hat{\ell}_z \!=\! 0$, while the condition for minimizing the Zeeman energy~\cite{brinkman}, $\hat{\ell}_z \!=\! 1$, is same as the condition that opens the maximum energy gap in the SABS. Hence, in the magnetic field lower than the dipolar field $H_{\rm D} \!\approx\! 30 {\rm G} \!\sim\! 0.001\pi T_{\rm c0}/\mu _{\rm n}$, as displayed in Fig.~\ref{fig:n}(a), $\hat{\bm n}$ points to $\hat{\bm z}$. It is seen in Fig.~\ref{fig:n}(b) that for the larger $H$'s it tends to tilt from $\hat{\bm z}$ to the direction with $\hat{\ell}_z \!\neq\! 0$. 

The field dependence of $\hat{\ell}_z$ estimated with the stable
configuration of $(\hat{\bm n},\varphi)$ is displayed in
Fig.~\ref{fig:n}(d). In the limit of the low field, $\hat{\ell}_z$ is
locked to be $\hat{\ell}_z \!=\! 0$, which ensures the existence of
surface Majorana fermions. $\hat{\ell}_z$ stays zero up to the critical
value $\mu _{\rm n} H^{\ast}/\pi T_{\rm c0} \!\approx\! 0.001$, which is
consistent with the argument that the systems with $\hat{\ell}_z \!=\!
0$ has the discrete symmetry. At $H \!\ge\! H^{\ast}$, the symmetry
protected topological phase with $\hat{\ell}_z \!=\! 0$ undergoes a
change to non-topological phase with $\hat{\ell}_z \!\neq\! 0$.

\begin{figure}[t!]
\includegraphics[width=85mm]{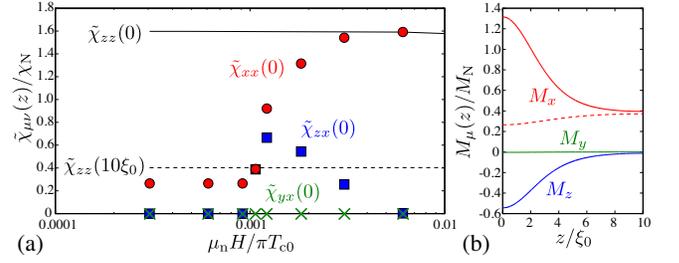}
\caption{(Color online) (a) Field dependence of $\tilde{\chi}_{\mu \nu}(z)/\chi _{\rm N}$ at $T \!=\! 0.2T_{\rm c0}$. The solid (dashed) lines denote
$\tilde{\chi}_{zz}(0)$ ($\tilde{\chi}_{zz}(10\xi)$) for ${\bm H}\!\parallel\! \hat{\bm z}$ and the symbols correspond to $\tilde{\chi}_{\mu x}(0)$ for ${\bm H}\!\parallel\! \hat{\bm x}$. (b) $M_{\mu}(z)$ for ${\bm H} \!\parallel\! \hat{\bm x}$ at $\mu _{\rm n} H /\pi T_{\rm c0} \!=\! 9.2 \!\times\!10^{-4}$ (dashed line) and $0.0018$ (solid lines), where $M_{y,z}$ at $\mu _{\rm n} H /\pi T_{\rm c0} \!=\! 9.2 \!\times\!10^{-4}$ are zero. All data are taken with $\tilde{\Lambda}_{\rm D}/\Lambda^2 \!=\! 2 \!\times\! 10^{-4}$.}
\label{fig:mag}
\end{figure}

In Fig.~\ref{fig:mag}(a) we plot the field dependence of the local spin susceptibility on the surface, $\tilde{\chi}_{\mu z}(z \!=\! 0)$, defined as $\tilde{\chi}_{\mu \nu}(z)/\chi _{\rm N} \!\equiv\! M_{\mu}(z)/M_{\rm N}$ for a magnetic field ${\bm H} \!\parallel\! \hat{\bm r}_{\nu}$. The local magnetization $M_{\mu}(z)$ is estimated as  
$
M_{\mu} ({\bm r}) \!=\! M_{\rm N} [ \hat{h}_{\mu} + \frac{1}{\mu _{\rm n}H}
\langle g_{\mu}(\hat{\bm k},{\bm r};i\omega _m)
\rangle_{\hat{\bm k},\omega_m}  ]
$
with that in the normal state $M_{\rm N} \!=\! \frac{2\mu^2_{\rm n}}{1+F^{\rm a}_0} N_{\rm F} H$, where $N_{\rm F}$ is the density of states of the normal $^3$He. It is seen from Fig.~\ref{fig:mag}(a) with the solid line that for ${\bm H} \!\parallel\! \hat{\bm z}$, the local spin susceptibility on the surface, $\tilde{\chi}_{zz}(0)$, is considerably enhanced, compared with $\tilde{\chi}_{zz}(z\!=\!10\xi )$ (the dashed line) \cite{TMfull}. 

In contrast, when the parallel field (${\bm H} \!\parallel\! \hat{\bm x}$) is applied, the magnetization $M_{\mu}(z)$ on the surface is sensitive to the orientation of $\hat{\bm \ell}$. It is seen in Fig.~\ref{fig:mag}(b) with the dashed line that $M_x(z)$ at $\mu _{\rm n} H /\pi T_{\rm c0} \!=\! 9.2 \!\times\!10^{-4}$ is strongly suppressed in the surface region, where $\hat{\bm n}\!\parallel\! \hat{\bm z}$, that is $\hat{\ell}_z \!=\! 0$, is energetically favored. This implies that the SABS does not contribute to $M_x(z)$ on the surface and is consistent with the property of the Majorana Ising spins. 

In the relatively high field $\mu _{\rm n}H/\pi T_{\rm c0}\!=\! 0.0018$, however, $M_x(z)$ is enhanced around the surface, while $M_z(z)$ which is perpendicular to ${\bm H}\!\parallel\! \hat{\bm x}$ emerges on the surface. This emergence of $M_z(z)$ on the surface reflects the stable configuration of $(\hat{\bm n},\varphi)$, where $\hat{\ell}_z \!=\! R_{xz}(\hat{\bm n},\varphi)$ deviates from zero and is less than unity. As displayed in Figs.~\ref{fig:mag}(a) and \ref{fig:mag}(b), the magnetic field within the range of $0\!<\!\hat{\ell }_z \!<\! 1$ significantly induces $M_{z}(z)$ and $\tilde{\chi}_{zx}(z)$ on the surface, where the SABS opens the finite energy gap and the winding number $w$ is not defined.

{\it Conclusions.---}
Here, we have clarified that the interplay of dipole interaction and a magnetic field in $^3$He-B involves a new class of quantum phase transition, that is, the topological phase transition with the spontaneous breaking of the hidden ${\bm Z}_2$ symmetry. Using the quasiclassical theory, we have demonstrated that $^3$He-B stays topological as the symmetry protected topological order and Majorana Ising spins exist unless the ${\bm Z}_2$ symmetry is spontaneously broken at the critical field $H^{\ast}$. The quantum phase transition is accompanied by the anomalous behavior of spin susceptibilities, which is observable through NMR experiments in a slab geometry.

The authors are grateful to S. Higashitani, M. Ichioka, O. Ishikawa,
R. Nomura, J. Saunders, R. Shindou, Y. Okuda, and Y. Tsutsumi for
fruitful discussions and comments. This work was supported by JSPS
(No.~2074023303, 2134010303 and 22540383) and the MEXT KAKENHI (No.~22103002 and
No.~22103005).  


\end{document}